# On-line Assembling Mitochondrial DNA from de novo transcriptome

Santiago Passos Patiño
Universidad Eafit
Colombia
spassos@eafit.edu.co

Juan David Arcila Moreno
Universidad Eafit
Colombia
jarcil13@eafit.edu.co

Mauricio Toro
Universidad Eafit
Colombia
mtorobe@eafit.edu.co

## ABSTRACT
This paper is focused in designing an efficient on-line algorithm to reconstruct a DNA sequence and search the genes in it, we assume that the segment have no mutation or reading error, the algorithm is based on de Bruijn Graph for reconstructing the DNA from the segments taking k-mers large enough no to generate cycles, once the sequence is ready a Boyer-Moore's algorithm implementation is used to search the genes inside de sequence using starts and stop codons, this solution give a high performance when all genes can be found, and there is no need to read all the segments to reach maximum number of genes, but due to the online nature one cannot be sure about the finals genes given.

## PROBLEM
We are facing DNA assembling and gen recognition, from some DNA segments, the original sequence must be fully reconstructed, and then find the genes that it contains. It must be developed an on-line algorithm that will constantly reading DNA sequences and finding the genes inside them, this process must be done efficiently in terms of memory and time with the less possible mistakes generated in the final complete sequence.

## INTRODUCTION
Today DNA sequencing algorithm do not use a genome reference to reconstruct the sequence, actual technologies like Stanford nano-pore read a lot of segments from a cell, this segment must be reconstructed in the original DNA, then search the genes to identify what specie it belongs to. To solve this problem an on-line algorithm is used to reconstruct and search the genes in the sequence, all while it's reading more segments.

## 1. Related Work

### [1] A de novo Genome Assembler based on MapReduce and Bi-directed De Bruijn Graph

This paper talk about the development of an algorithm that uses MapReduce and Bi-directed De Bruijn Graph for efficient DNA assembly, MapReduce is used in the process to parallelize and optimize the construction, compaction and cleaning of the graph.

First the algorithm read the sequences from the de novo, and while reading, it creates an undirected De Bruijn

Graph, this kind of graph in its classical way is a directed graph, because of this, creating a Bi-directed form require some specials changes during the construction, after that it must be cleaned, because during the reading it may happen some errors, so the algorithm eliminated duplicated, and unnecessary nodes, and repeated or useless paths between the nodes. The pros of this implementation are that the traversal of the graph once it's already cleaned (fixed all the bubbles and gaps) it's very easy to reconstruct the genome, giving more consistent results with less errors.

### [2] HipMer: An Extreme-Scale De Novo Genome Assembler

In this article, the authors explore efficient ways of: generating the k-mers of the gnome, traverse the De Bruijn graph and solve the errors occurred during the DNA reading.

For the first step while reading the sequence they categorize the k-mers according of its occurrence in the sequence, that way the ones with less recurrence can be determined as an error, as this algorithm is applied the memory used is reduced by 85%. Another improvement is in the traversal of the De Bruijn graph where the algorithm divide the graph into different contigs, so each of the contigs can be traversed by a different processor dividing the amount of time required for reconstructing the genome. The last step of the algorithm is to identify and solve the problems caused by the bad reading, things like overlapped contigs, gaps between contigs, and ambiguous contigs of the graph called bubbles, this is solved by parallelizing the process and generating different states of the genomes with different possibilities of reconstructing it.

### [3] Using Matching DNA sequences Algorithms: Aho-Corasick (AC) and Boyer Moore (BM)

Is usual to see these two algorithms to make high-efficiency comparison between two or more completes DNA. In our actual application, we search two equal n-sequence (Prefix and Suffix) present in two segments of DNA, because those algorithms were made for exact string matching and multi pattern finding, and we are looking for a not exactly length string, however, it can be very useful to search the longest n-substring common in all sequences, something that we need to construct K-Overlap graph or De Bruijn Graph. For this reason, is important to consider in our solutions this kind of Algorithms.

The Aho-Corasick algorithm use two main stages: A Finite state machine construction stage and a matching stage. The finite state machine construct a suffix-tree, and the matching stage makes find out the pattern set within the given string. This make a "graph" that can contain cycles, and make possible find all the paths given the suffix of the sequences and patterns.

The Boyer Moore algorithm is an efficient string searching algorithm that is consider
the most efficient string-matching algorithm in usual applications, for example, in text editors and commands substitutions. The reason is that it works the fastest when the alphabet is moderately sized (in this case 4) and the pattern is relatively long.

[4] K-Overlap Graph with TSP

A K-overlap graph is a di-graph in which each string in a collection is represented by a node and node *s* is connected to *t* with a directed edge if and only if some suffix of *s* equals a prefix of *t*. We say that *k* is the length of the suffix and prefix present in both strings, the weight of every edge is the k for those nodes connected by this edge. For our application, we assume that the length of the string is more than *k* and less than 100, also the nodes are the n-segments of DNA. To see the final and complete DNA sequence we find the shortest path which visits every node exactly once, this is the TSP (Traveling Salesman Problem).

The problem with this solution is that we are using NP-complete problems to find the complete DNA sequence, more exactly the TSP and get the common k-substrings between all the n-segments, so its complexity is not good, but always give a correct solution. Another important characteristic of this solution is that it can't consider the possible ambiguity between the overlap of two or more segments, in that case we need to make a check to verify if every three nucleobases of the ARNm exist, in that case, the DNA sequence is correct.

**2. Data Structure Selection**

Within all the algorithms to choose, we opt for the Bruijn graph due to its ease for reconstruction.

| Solution | Time Complexity | Memory Complexity |
|---|---|---|
| Brute force | O(n!) | O(n!) |
| K-overlap | O (2^n * n^2) | O (2^n * n) |
| Bruijn Graph | O (nm + e) | O(e ) |

**Table 1:** Solution comparison
n = number of segments to read
m = time to get a segment into k-mers
e = time to traverse the graph

The data structure selected was De Bruijn Graphs, with this the sequence can be reconstructed, using K-mers for construct every node in the graph. Every time that a string is read the program divide the sequence into K-mers to construct the nodes and the edges for the graph.

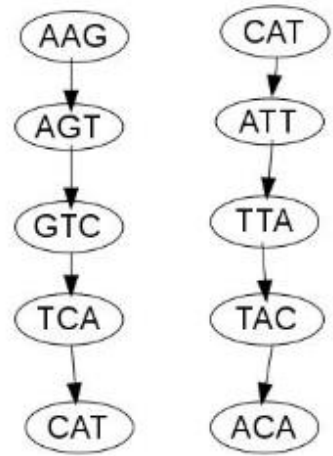

**Graph 1:** Example of sequence divided in k-mers

Once the graph is completely constructed the graph is traversed in order, from the first node (the one that is not receiving an edge) to its adjacent node, this process continue until the last node is reached, then the original DNA sequence is finally reconstructed.

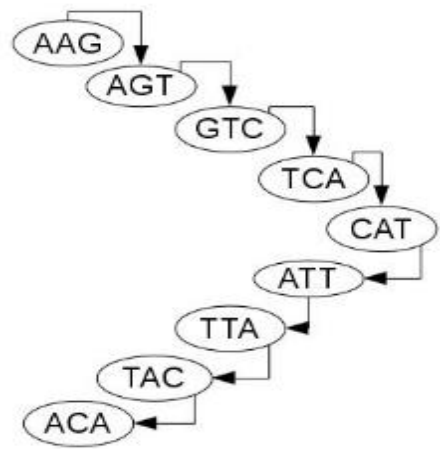

**Graph 2:** Example of de Bruijn Graph reconstruction
This graph represents the sequence AAGTCATTACA

## 3. Algorithm Design

Using the De Bruijn Graph a DNA sequence can be reconstructed, then the genes inside must be sought, for this the Boyer-Moore's Algorithm is used to search the index of the start (AUG) and stop codons (UAA, UAG, UGA), then it takes the left most possible start codon, and try to find a possible gen with the stop codons, once it finds a gen it takes the next start codon which its index is greater than the last gen stop codon, this process continue until no more genes can be found.

The program prints every gen that find, because every time new segments are read, the genes found can change during the process, this until there are no more segments then, the final gen printed are the genes in the original sequence.

## 4. Complexity

The complexity of the algorithm is divided in three parts, the first one is the construction of the graph, in this case the algorithm read N strings to reconstruct the DNA sequence, each of this string are divided into k-mers. This require n the length of the read, then encode each k-mer in a hash table, that action is executed in O (1), the total complexity is O(Nm), m is the time required to separate the segments into k-mers that give us a complexity of O(Nm). The second part of the complexity is the reconstruction of the sequence, that is given by the traverse of the graph, in this case, if one pick enough large K-mer the graph is like a straight line, then the time to go through all the graph is |E| the numbers of edges, this is the same as S-1, so it's a linear time O(S) where S is the maximum number of k-mers.

The last part is the search of the genes, for this is used Boyer-Moore's algorithm which search the index in O(n/m) where n is the size of the string and m is the size of the pattern that want to be find, then to generate all the genes is required O(s) where s is the number of start codons in the sequence, the final complexity is O (n/m + s).

The execution time was calculated by running the program with five different sequences each one with different number of gens (10, 20, 30, 40, 50).

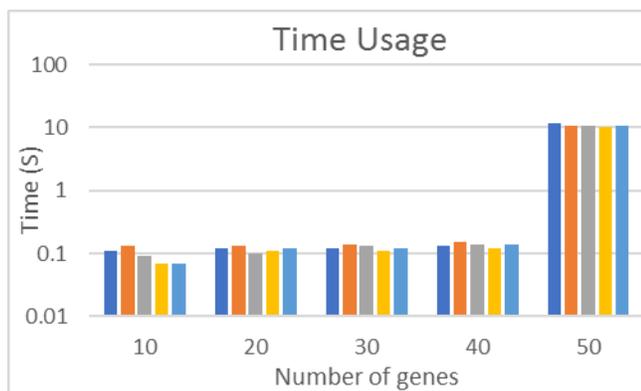

**Graph 3**: Comparison in the time used

The drastic difference in time requires is because when it looks for 50 genes the sequence doesn't have that amount of gen, then the program have to read all the segments.

The complexity in memory in the worst case is just the necessary to keep the map with the k-mers and the final DNA sequence, that give us O(S), the next table shows the memory required in different cases.

| Sequence Size | Memory (MB) |
|---|---|
| 16692 | 12.1 |
| 15600 | 11.6 |
| 16508 | 12 |
| 13794 | 10.3 |
| 17085 | 13.3 |

**Table 2**: Comparison in the memory used

## 6. Conclusions

After analyzing the different solutions for the problem, we conclude that the de Bruijn Graph is an optimal option to reconstruct the sequence due to its ease to traverse and its integrity about maintaining the original DNA string, and Boyer-Moore offers a quickly way to find codons inside the DNA, in a way that area easily processed to be defined as a gen or no, this on-line solution is viable because sequence with mutation or reading errors were not considered.

## ACKNOWLEDGMENTS
We thank or classmates Juan Manuel Ciro and Juan José Suarez for their useful and interesting discussions about the solutions, and thanks the teachers Mauricio Toro and Javier Correa for their guidance through all the work.